\documentclass[preprint,prd,showpacs,showkeys,preprintnumbers,aps]{revtex4}
\usepackage{graphicx}
\usepackage{dcolumn}

\def\ben{\begin{equation}}
\def\een{\end{equation}}
\def\bey{\begin{eqnarray}}
\def\eey{\end{eqnarray}}
\def\ba{\begin{array}}
\def\ea{\end{array}}
\def\benmrt{\begin{enumerate}}
\def\eenmrt{\end{enumerate}}

\def\psla{p{\raise1pt\hbox{$\!\!/$}}}
\def\dsla{\partial{\raise1pt\hbox{$\!\!\!/$}}}
\def\Dsla{D{\raise1pt\hbox{$\!\!\!/$}}}
\def\xsla{x{\raise1pt\hbox{$\!\!\!/$}}}

\def\jmu5{j_{\mu 5}^{(i)}(0)}
\def\jnu5{j_{\nu 5}^{(i)}(0)}

\def\qq0v{\langle0\!\mid\!{\bar q}q\!\mid\! 0\rangle}

\def\qc0f{\langle0\!\mid\!{\bar q}q\!\mid\!0\rangle_{F}}

\def\qsq0f{\langle0\!\mid\!{\bar q}\sigma_{\mu\nu}q\!\mid\!0\rangle_{F}}

\def\qgdq0f{\langle0\!\mid\!{\bar q}{\cal 
S}\gamma_{\mu}D_{\nu}q\!\mid\!0\rangle_{F}}

\def\qddq0f{\langle0\!\mid\!{\bar q}{\cal
S}D_{\mu}D_{\nu}q\!\mid\!0\rangle_{F}}

\def\3mmtm{|{\bf q}|^2}

\def\eq#1{Eq.(\ref{#1})}
\def\eqs#1#2{Eqs.(\ref{#1}) and (\ref{#2})}

\def\meq#1#2{Eqs.(\ref{#1})$\sim$(\ref{#2})}

\def\Ref#1{[\ref{#1}]}
\def\Refs#1#2{[\ref{#1},\ref{#2}]}

\def\p0{p_0}

\def\gam3{\mbox{\boldmath{$\gamma$}}}
\def\e0{E_{0}(s_{0},s)}
\def\e1{E_{1}(s_{0},s)}
\def\e2{E_{2}(s_{0},s)}

\def\aplt{\kern0.3333em \raise 0.2ex \hbox{$<$}%
\kern-0.8em \lower0.8ex \hbox{$\sim$}%
\kern0.3333em}
\def\aplg{\kern0.3333em \raise 0.2ex \hbox{$>$}%
\kern-0.8em \lower0.8ex \hbox{$\sim$}%
\kern0.3333em}

  \def\S(#1,#2){{S^{#1}_{#2}}}
 \def\Su(#1,#2){{\hat{S}^{#1}_{#2}}}
 \def\Sd(#1,#2){{\check{S}^{#1}_{#2}}}
 \def\C(#1){C_{#1}}
  \def\CG(#1){(C\gamma_5)_{#1}}
  \def\G(#1){{\Gamma}^{#1}}
  \def\g{{\gamma}}  
  \def\q{{\bar q}}

  \def\s{{\bar s}}

   \def\Ci{C^{-1}}

  \def\gt_#1{(-C\g_{#1}\Ci)}
  \def\bra{\langle}
  \def\ket{\rangle}
  \def\qcond{\bra 0|\q q|0\ket}

  \def\sc{\bra 0|\s s|0\ket}

  \def\psla{p{\raise1pt\hbox{$\!\!/$}}}
  \def\qsla{q{\raise1pt\hbox{$\!\!\!/$}}}
  \def\knd(#1,#2){\delta_{#1#2}}

  \def\sGs{g\langle 0|\bar s\sigma^{\mu\nu}(\lambda^a/2)G^a_{\mu\nu}s|0\rangle}

  \def\GG{\langle 0|(\alpha_s/\pi)G^{a\mu\nu}G^{a}_{\mu\nu}|0\rangle}
  
  \def\vecp{\textnormal{\mathversion{bold}$p$}}
  \def\boldp{{\boldmath$p$}}

\begin{document}

\preprint{KEK-TH-997}

\title{Spin-3/2 pentaquark in the QCD sum rule}

\author{Tetsuo NISHIKAWA}
\email{nishi@post.kek.jp}
\author{Yoshiko KANADA-EN'YO}
\email{yoshiko.enyo@kek.jp} 
\affiliation{%
Institute of Particle and Nuclear Studies, 
High Energy Accelerator Research Organization, 1-1, Ooho, 
Tsukuba, Ibaraki, 305-0801, Japan
}
\author{Yoshihiko KONDO}
\email{kondo@kokugakuin.ac.jp}
\affiliation{%
Kokugakuin University, Higashi, Shibuya, Tokyo 150-8440, Japan
} 
\author{Osamu MORIMATSU}
\email{osamu.morimatsu@kek.jp}
\affiliation{%
Institute of Particle and Nuclear Studies, 
High Energy Accelerator Research Organization, 1-1, Ooho, 
Tsukuba, Ibaraki, 305-0801, Japan
}

\date{\today}
\begin{abstract}
We study $IJ^P=0\frac{3}{2}^\pm$ and $1\frac{3}{2}^\pm$ pentaquark states with $S=+1$ in the QCD sum rule approach. 
The QCD sum rule for positive parity states and that for negative parity are independently derived. The sum rule suggests that there exist the $0\frac{3}{2}^-$ and the $1\frac{3}{2}^-$ states.
These states may be observed as extremely narrow peaks since they
can be much below the $S$-wave threshold and since the only allowed decay channels are $NK$ in $D$-wave, whose centrifugal barriers are so large that the widths are strongly suppressed. 
The $0\frac{3}{2}^-$ state may be assigned to the observed $\Theta^+(1540)$ and the $1\frac{3}{2}^-$ state can be a candidate for $\Theta^{++}$.
\end{abstract}
\maketitle

\section{Introduction}
Recent observation of an exotic baryon state with positive strangeness, $\Theta^{+}(1540)$, by LEPS collaboration in Spring-8 \cite{leps} and subsequent experiments \cite{diana,clasa,clasb,saphir,itep,hermes,itep-2,zeus,clas-c} has raised great interests in hadron physics.
This state cannot be an ordinary three-quark baryon since having positive strangeness, and the minimal quark content is $(uudd{\bar s})$.
A remarkable feature of $\Theta^{+}(1540)$ is that the width is unusually small ($\Gamma<25\,{\rm MeV}$) despite the fact that it lies 
about $100$ MeV above the $NK$ threshold.
The absense of isospin partners suggests that the $\Theta^+$ is an isosinglet \cite{clasa,saphir}. The spin and parity have not yet been experimentally determined.

The discovery of $\Theta^{+}$ has triggered intense theoretical studies
to understand the structure of the $\Theta^{+}$ \cite{jaffe,karliner,sasaki,csikor,zhu,matheus,sugiyama,eide,amdpenta,hosaka,takeuchi}.
One of the main issues is to clarify the quantum numbers, especially, the spin and the parity, which are key properties for understanding 
the abnormally small width.

Following a naive expectation from ordinary hadron spectra, 
it is natural to assume that the $\Theta^{+}$ has the lowest spin $J=\frac{1}{2}$.
In fact, most of the existing works on
$\Theta^{+}$ in lattice QCD \cite{sasaki,csikor} or in QCD sum rule \cite{zhu,matheus,sugiyama,eide} have focussed only on the $J=\frac{1}{2}$ states.
However, we cannot exclude the possibility that $\Theta^{+}$ is a higher spin state, as suggested in some literatures \cite{close,amdpenta,hosaka,takeuchi,jaffe2}.
For example, negative parity $J=\frac{3}{2}$ states are 
especially important because they can be extremely narrow
in the following reasons:
First, consider an $IJ^P=0\frac{3}{2}^-$ state ($I$ and $P$ denote the total isospin and parity, respectively).
If this state lies much below the $NK^*$ threshold, 
no $S$-wave decay channel opens and 
the decay is restricted only to $D$-wave $NK$ states.
Due to the high centrifugal barrier the width is strongly suppressed.
Thus, this state can be a candidate for the observed $\Theta^{+}$.
For just the same reason, $1\frac{3}{2}^-$ can also be 
seen as a narrow peak.
If the state is sufficiently below the $\Delta K$ threshold, the allowed decay channel is only $NK$ in $D$-wave and the width can be significantly small
\cite{amdpenta,theta++}.

Another state in which we have an interest is the pentaquark with 
$IJ^P=0\frac{3}{2}^+$, which has been discussed as a spin-orbit ($LS$) partner of the $0\frac{1}{2}^+$ state \cite{close}.
The $0\frac{1}{2}^+$ state was assigned to the observed $\Theta^+$ by Jaffe and Wilczek \cite{jaffe}.
According to their conjecture,  the $\Theta^+$ consists of two spinless $ud$-diquarks in $P$-wave and an anti-strange quark.
As a result, the $LS$-partners, $0\frac{1}{2}^+$ and $0\frac{3}{2}^+$,
appear due to the coupling between the relative motion with orbital angular momentum $L=1$ and the intrinsic spin of the anti-strange quark. The effect of the $LS$ force is so weak in the diquark structure that the mass splitting should be small. It leads to a possibility of the low-lying $\Theta^*(IJ^P=0\frac{3}{2}^+)$  
as pointed out in Ref.\cite{close}.

In this paper, we study the $0\frac{3}{2}^\pm$ and $1\frac{3}{2}^\pm$ pentaquark states by using the method of QCD sum rule \cite{SVZ}.
In order to ascertain the existence of the $0\frac{3}{2}^-$ and $1\frac{3}{2}^-$ narrow pentaquarks, it is crucial to estimate their absolute masses
since their widths are sensitive to the energy difference from the $NK^*$ or $\Delta K$ threshold.  
QCD sum rule is closely related to the fundamental theory and is able to evaluate the absolute masses of hadrons without any model assumptions.
In QCD sum rule approach, a correlation function of an interpolating field is calculated by the use of 
the operator product expansion (OPE),
and is compared with the spectral representation via dispersion relation.
The sum rules relate hadron properties to the vacuum expectation values
of QCD composite operators (condensates) such as $\langle 0|{\bar q}q|0\rangle$, $\langle 0|(\alpha_{s}/\pi)G^2|0\rangle$ and so on.
From the relation, one can understand hadron properties in terms of the structure of the QCD vacuum.

The paper is organized as follows. In the second section, we formulate the general method for deriving the QCD sum rules for positive and negative parity baryons with $J=3/2$. Then, in the third section, we apply the method to constructing the sum rules for the pentaquark with $IJ^P=0\frac{3}{2}^\pm$ and that with $1\frac{3}{2}^\pm$.
From the obtained sum rules, we show the numerical results, discuss whether those states of pentaquark can exist, and evaluate their masses in the fourth section.
In the fifth section, we discuss whether they can be narrow states and
the relation with the investigation by other approaches,
and finally summarize the paper.

\section{QCD sum rule for positive and negative parity baryons with $J=3/2$}

In this section, we formulate the QCD sum rule for positive and negative parity states of $J=3/2$ baryons.

The correlation function from which we derive the QCD sum rule is
\bey
\Pi_{\mu\nu}(p)&=&-i\int d^4 x\exp(ipx)
\langle 0|T\left[\eta_{\mu}(x)\overline\eta_{\nu}(0)\right]|0\rangle,
\label{corfun}
\eey
where ${\eta}_{\mu}$ is an interpolating field that couples to baryon states with $J=3/2$.
${\eta}_{\mu}$ is constructed with quark (and gluon) fields so as to have the quantum numbers of the baryon which we want to know about.
The correlation function, \eq{corfun}, has various tensor structures,
\bey
\Pi_{\mu\nu}(p)&=&\left[\Pi_1(p^2)g_{\mu\nu}\psla+\Pi_2(p^2)\g_\mu \g_\nu\psla
+\Pi_3(p^2)\g_\mu p_\nu+\Pi_4(p^2)\g_\nu p_\mu+\Pi_5(p^2)p_\mu p_\nu \psla\right]\cr
&&+\left[\Pi_6(p^2)g_{\mu\nu}+\Pi_7(p^2)\g_\mu \g_\nu
+\Pi_8(p^2)\g_\mu p_\nu\psla+\Pi_9(p^2)\g_\nu p_\mu\psla+\Pi_{10}(p^2)p_\mu p_\nu \right].
\eey
We consider the terms proportional to $g_{\mu\nu}$:
\bey
\Pi(p)&\equiv&\psla\Pi_{1}(p^2)+\Pi_{6}(p^2),
\label{3/2part}
\eey
since only the $J=3/2$ states contribute to these terms.
Other terms include the contribution from not only $J=3/2$ states but also $J=1/2$ states \cite{ioffe}.

We can relate the correlation function with the spectral function via Lehman representation, 
\begin{eqnarray}\label{Lehman}
\Pi(p_0,\vecp
)&=&\int_{-\infty}^{\infty}{\rho(p'_0,\vecp)\over p_0-p'_0}dp'_0,
\end{eqnarray}
where $\rho(p_0,\vecp)\equiv(1/\pi){\rm Im}\Pi(p_0+i\epsilon,\vecp
)$ is the spectral function.
On the other hand, in the deep Euclid region, $p_0^2\rightarrow-\infty$, 
the correlation function can be evaluated by an OPE. Then the correlation function is expressed as a sum of various vacuum condensates.
Using the analyticity, we obtain a relation between the imaginary part of the correlation function evaluated by an OPE, $\rho^{\rm OPE}(p_0,\vecp)$, and the spectral function as
\begin{eqnarray}
\int_{-\infty}^{\infty}dp_0\rho^{\rm OPE}(p_0,\vecp)W(p_0)
&=&\int_{-\infty}^{\infty}dp_0\rho(p_0,\vecp)W(p_0),
\label{QSR}
\end{eqnarray}
where $W(p_0)$ is an analytic function of $p_0$.
\eq{QSR} is a general form of the QCD sum rule.
By properly parameterizing $\rho(p_0,\vecp)$, we obtain
QCD sum rules for physical quantities in $\rho(p_0,\vecp)$.

Let us first look at the spectral function, $\rho(p_0,\vecp)$.
We consider the interpolating field with negative parity.
The interpolating field couples to positive parity states as well as negative parity states \cite{chung}.
Accordingly, both of the parity states contribute to $\Pi(p)$ as physical intermediate states. 
The expression of $\Pi(p)$ in terms of the physical states reads \cite{ioffe},
\begin{eqnarray}
\Pi(p)&=&\sum_n\left[-|\lambda^{(n)}_-|^2{\psla+m^{(n)}_-\over p^2-{m^{(n)}_-}^2}
-|\lambda^{(n)}_+|^2{\psla-m^{(n)}_+\over p^2-{m^{(n)}_+}^2}\right],
\label{pi-spe}
\end{eqnarray}
where $m^{(n)}_{+,-}$ are the masses of the $n$-th positive and negative parity states, respectively, and $\lambda^{(n)}_{+,-}$ the coupling strengths of the interpolating field with the positive and negative parity states.
In \eq{pi-spe}, the widths of the physical states were neglected.
The spectral function in the rest frame, $\vecp={\bf 0}$, can be decomposed into two parts,
\begin{eqnarray}
\rho(p_0)&=&
P_+\rho_{-}(p_0)+P_-\rho_{+}(p_0),
\label{Spectral-negativeeta}
\end{eqnarray}
where $P_\pm=(\g_0\pm1)/2$ and $\rho_{\mp}(p_0)$ are expressed as
\bey
\rho_{-}(p_0)&=&\sum_n\left[|\lambda^{(n)}_-|^2\delta(p_0-m^{(n)}_-)|+|\lambda^{(n)}_+|^2\delta(p_0+m^{(n)}_+)\right],
\label{spe-}\\
\rho_{+}(p_0)&=&\sum_n\left[|\lambda^{(n)}_+|^2\delta(p_0-m^{(n)}_+)+|\lambda^{(n)}_-|^2\delta(p_0+m^{(n)}_-)\right].
\label{spe+}
\eey
Here, for later use, we note that the spectral function for the interpolating field with positive parity is given by interchanging $\rho_{-}(p_0)$ and $\rho_{+}(p_0)$ in \eq{Spectral-negativeeta},
\begin{eqnarray}
\rho(p_0)&=&
P_+\rho_{+}(p_0)+P_-\rho_{-}(p_0),\quad(\mbox{
for $\eta_{\mu}$ with positive parity }).
\label{Spectral-positiveeta}
\end{eqnarray}

Next, we construct the sum rule for negative parity states and that for positive parity. We apply the projection operator $P_\pm$ to \eq{QSR} for $\vecp={\bf 0}$. Then we obtain
\begin{eqnarray}
\int_{-\infty}^{\infty}dp_0\rho_{\mp}^{\rm OPE}(p_0)W(p_0)
&=&\int_{-\infty}^{\infty}dp_0\rho_{\mp}(p_0)W(p_0).
\label{projectQSR}
\end{eqnarray}
Note that in \eq{projectQSR} the contributions from the positive and negative parity states are mixed
since, as can be seen from \eqs{spe-}{spe+}, each of $\rho_{-}(p_0)$ and  $\rho_{+}(p_0)$ contains the contributions from both of the parity states.
What we want to do is to separate the negative and the positive parity contributions from Eqs.(\ref{projectQSR}).
The following procedure of the parity projection is essentially equivalent to that in Ref.\cite{jido}.

If the expressions of the spectral functions calculated by an OPE, $\rho_{\mp}^{\rm OPE}(p_0)$, are separable into $\rho_{\mp}^{\rm OPE}(p_0>0)$ and $\rho_{\mp}^{\rm OPE}(p_0<0)$, 
we can independently construct the sum rule from the $p_0>0$ part of the correlation function and that from the $p_0<0$ part \cite{jido}.
The sum rules obtained from the $p_0>0$ part are
\begin{eqnarray}
&&\int_{0}^{\infty}dp_0\rho_{-}^{\rm OPE}(p_0)W(p_0)
=\int_{0}^{\infty}dp_0\rho_{-}(p_0)W(p_0),
\label{compprojectQSR-}\\
&&\int_{0}^{\infty}dp_0\rho_{+}^{\rm OPE}(p_0)W(p_0)
=\int_{0}^{\infty}dp_0\rho_{+}(p_0)W(p_0).
\label{compprojectQSR+}
\end{eqnarray}
\eq{compprojectQSR-} is the sum rule for the negative parity state
since only the negative parity states contribute to $\rho_{-}(p_0)$ for $p_0>0$.
On the other hand, \eq{compprojectQSR+} is the sum rule for the positive parity state since $\rho_{+}(p_0)$ for $p_0>0$ contains only the positive parity states
(see \eqs{spe-}{spe+}).

A comment is in order here.
In order to separate the positive and negative parity states in the sum rule as \eqs{compprojectQSR-}{compprojectQSR+},
it is necessary that $\rho_{\mp}^{\rm OPE}(p_0)$ are separable into $\rho_{\mp}^{\rm OPE}(p_0>0)$ and $\rho_{\mp}^{\rm OPE}(p_0<0)$
as mentioned above.
In general, $\rho_{\mp}^{\rm OPE}(p_0)$ are not separable \cite{kondo} because the OPE terms depend on $(p_0)^n\left[\theta(p_0)-\theta(-p_0)\right]$ or $\delta^{(n)}(p_0)$.
However, as will be seen in the next section, $\rho_{\mp}^{\rm OPE}(p_0)$
for pentaquark is separable as long as we truncate the OPE at certain order
since $\rho_{\mp}^{\rm OPE}(p_0)$ up to dimension 7 operator contain
only the $(p_0)^n\left[\theta(p_0)-\theta(-p_0)\right]$ terms.
We can thus derive the sum rule for each parity state of the pentaquark as \eqs{compprojectQSR-}{compprojectQSR+}.

In \eqs{compprojectQSR-}{compprojectQSR+}, we parameterize $\rho_\mp(p_0)$ for $p_0>0$ with a pole plus continuum contribution,
\begin{eqnarray}\label{Phen}
\rho_\mp(p_0)&=&|\lambda_\mp|^2\delta(p_0-m_\mp)+\theta(p_0-\omega_\mp)\rho^{\rm OPE}(p_0),\quad
\mbox{for $p_0>0$},
\end{eqnarray}
where $|\lambda_\mp|^2$ and $m_\mp$ are the pole residues and the masses of the lowest states, respectively. $\omega_\mp$ denote the effective continuum threshold. 
Substituting \eq{Phen} into the right-hand sides of \eqs{compprojectQSR-}{compprojectQSR+},
we obtain the following sum rules,
\begin{eqnarray}\label{BSR}
\int_{0}^{\omega_\mp}dp_0\rho_\mp^{\rm OPE}(p_0)(p_0)^n\exp(-{p_0^2\over M^2})&=&(m_\mp)^n|\lambda_\mp|^2\exp(-{{m_\mp}^2\over M^2}).
\end{eqnarray}
Here we have chosen the weight function as $W(p_0)=(p_0)^n\exp(-p_0^2/M^2)$, where $n$ is an arbitrary positive integer.
The parameter $M$ is called \lq\lq Borel mass".
By introducing such weight function, one can improve the convergence of the OPE and simultaneously suppress the continuum contribution. 

From \eq{BSR} for $n=0$, we obtain the sum rule for the pole residues $|\lambda_\mp|^2$,
\bey
|\lambda_\mp|^2\exp(-{{m_\mp}^2\over M^2})&=&\int_{0}^{\omega_\mp}dp_0\rho_\mp^{\rm OPE}(p_0)\exp(-{p_0^2\over M^2}).
\label{residueSR}
\eey
The masses can be extracted from the ratio of \eq{BSR} for $n=0$ and $n=2$,
\bey
{m_\mp}^2=\frac{\int_{0}^{\omega_\mp}dp_0\rho_\mp^{\rm OPE}(p_0)(p_0)^2\exp(-{p_0^2\over M^2})}{\int_{0}^{\omega_\mp}dp_0\rho_\mp^{\rm OPE}(p_0)\exp(-{p_0^2\over M^2})}.
\label{massSR}
\eey

\section{QCD sum rules for the $IJ^P=0\frac{3}{2}^\pm$ and $1\frac{3}{2}^\pm$ states of pentaquark}
Utilizing the method formulated in the previous section,
we derive the QCD sum rules for the $IJ^P=0\frac{3}{2}^\pm$ and $1\frac{3}{2}^\pm$ states of the pentaquark baryons.

Our first task is to construct the interpolating fields for the spin-3/2 pentaquark baryons. There are various ways of constructing interpolating fields.
In this paper, we examine two independent interpolating fields for each isospin state.
Then, the correlation functions of the interpolating fields are evaluated by the use of OPE.
For each isospin state, we choose the interpolating field which has better convergence of OPE among the two, and construct the sum rules.

\subsection{Interpolating field}
The interpolating fields for $I=0$ state which we employ are those proposed by Sasaki \cite{sasaki},
\bey
{\eta}_{1,\mu}^{I=0}(x)&=&\epsilon_{cfg}
\left[\epsilon_{abc}u_{a}^{T}(x)C\gamma_5 d_{b}(x)\right]
\left[\epsilon_{def}u_{d}^{T}(x)C\gamma_{\mu}\g_5 d_{e(x)}\right]
C\bar{s}_{g}^{T}(x),
\label{currentisoscalarS}\\
{\eta}_{2,\mu}^{I=0}(x)&=&\epsilon_{cfg}
\left[\epsilon_{abc}u_{a}^{T}(x)C d_{b}(x)\right]
\left[\epsilon_{def}u_{d}^{T}(x)C\gamma_{\mu}\g_5 d_{e(x)}\right]
\gamma_5 C\bar{s}_{g}^{T}(x),
\label{currentisoscalarP}
\eey
where $u$, $d$ and $s$ are up, down and strange quark fields, resepectively, roman indices $a,b,\ldots$ are color, $C=i\gamma^2 \gamma^0$ is the charge conjugation matrix, and $T$ transpose.
\eq{currentisoscalarS} consists of two diquark fields,
$S_c(x)\equiv\epsilon_{abc}u_{a}^{T}(x)C\gamma_5 d_{b}(x)$ and
$V_f(x)\equiv\epsilon_{def}u_{d}^{T}(x)C\gamma_{\mu}\g_5 d_{e}(x)$,
and an anti-strange quark field, $C\bar{s}_{g}^{T}(x)$.
The color structure is $\bar{\bf 3}\otimes\bar{\bf 3}\otimes\bar{\bf 3}$.
$S_c(x)$ is a color $\bar{\bf 3}$ scalar diquark operator with $I=0$, 
which corresponds to the ${}^1S_{0}$ state of the $I=0$ $ud$-diquark system.
$V^f(x)$ is a color $\bar{\bf 3}$ vector diquark with $I=0$, and is assigned to ${}^3P_{1}$ of the $I=0$ $ud$-diquark system.
\eq{currentisoscalarS} is therefore totally $I=0$, and hence, one can confirm that \eq{currentisoscalarS} can create
the states with $IJ=0\frac{3}{2}$.
The parity of \eq{currentisoscalarS} is positive since the intrinsic parity of $C\bar{s}_{g}^{T}(x)$ is negative.

Alternatively, we can construct the interpolating field for $I=0$ by using a pseudo-scalar diquark operator, $P_c(x)\equiv\epsilon_{abc}u_{a}^{T}(x)C d_{b}(x)$, instead of the scalar diquark.
$P_c(x)$ corresponds to the ${}^3P_{0}$ state of the $I=0$ $ud$-diquark 
system. 
In \eq{currentisoscalarP}, we multiplied $C\bar{s}_{g}^{T}(x)$ by $\gamma_5$ to make the total parity positive.
Clearly, \eq{currentisoscalarS} can also couple with $IJ=0\frac{3}{2}$ states.

In a quite similar way, we can construct the following two interpolating fields 
for the $I=1$ states, using the scalar or pseudo-scalar diquark and an axial-vector diquark operator \cite{theta++},
\bey
{\eta}_{1,\mu}^{I=1}(x)&=&\epsilon_{cfg}
\left[\epsilon_{abc}u_{a}^{T}(x)C\gamma_5 d_{b}(x)\right]
\left[\epsilon_{def}u_{d}^{T}(x)C\gamma_{\mu}u_{e}(x)\right]
C\bar{s}_{g}^{T}(x),
\label{currentisovectorS}\\
{\eta}_{2,\mu}^{I=1}(x)&=&\epsilon_{cfg}
\left[\epsilon_{abc}u_{a}^{T}(x)C d_{b}(x)\right]
\left[\epsilon_{def}u_{d}^{T}(x)C\gamma_{\mu}u_{e}(x)\right]
\gamma_5 C\bar{s}_{g}^{T}(x),
\label{currentisovectorP}
\eey
where $A_f(x)\equiv\epsilon_{def}u_{d}^{T}(x)C\gamma_{\mu}u_{e}(x)$
is a color $\bar{\bf 3}$ axial-vector diquark with $I=1$.
$A_f(x)$ corresponds to ${}^3S_{1}$ of the $I=1$ $ud$-diquark system.
One can easily see that either of \eq{currentisovectorS} and \eq{currentisovectorP}
contains $IJ=1\frac{3}{2}$ states.
It should be noted that total parity of \eqs{currentisovectorS}{currentisovectorP} is negative.
We remark that the way of constructing \eq{currentisovectorS} is based on the structure of the $1\frac{3}{2}^-$ pentaquark state suggested from a quark model \cite{amdpenta}.

\subsection{OPE}
We now evaluate the correlation functions of the interpolating fields, \meq{currentisoscalarS}{currentisovectorP},
\bey
\Pi_{j,\mu\nu}^{I}(p)&=&-i\int d^4 x\exp(ipx)
\langle 0|T\left[\eta_{j,\mu}^{I}(x)\overline\eta_{j,\nu}^{I}(0)\right]|0\rangle,\cr
&=&g_{\mu\nu}\Pi_{j}^{I}(p)+\mbox{(other tensor structures)},\quad(I=0,1,\,\,j=1,2),
\label{corfunSP}
\eey
by the use of OPE.
We take into account the terms up to dimension 7 operators and neglect the masses of up and down quarks.
The spectral function of $\Pi_{j}^{I}(p)$ for \boldp=0,  $\rho_{j}^{I}(p_0)$, is parametrised in terms of the chirality conserving term and the violating term, which we denote by $A_{j}^{I}(p_0)$ and $B_{j}^{I}(p_0)$, respectively, as
\bey
\rho_{j}^{I}(p_0)&=&\g_0 A_{j}^{I}(p_0)+B_{j}^{I}(p_0).
\label{ope}
\eey
We show the results of the OPE of $A_{j}^{I}(p_0)$ and $B_{j}^{I}(p_0)$ in the following.

When we use the interpolating field ${\eta}_{1,\mu}^{I=0}(x)$, we obtain
\bey
A_1^{I=0}(p_0)&=&
\left[a_0\cdot (p_0)^{11}+a_1\cdot  \GG(p_0)^7
\right.\cr
&&
+a_2\cdot  m_{s}\sc(p_0)^{7}+a_3\cdot m_{s}\sGs(p_0)^{5}\cr
&&\left.
+a_4\cdot \qcond^2 (p_0)^{5}
\right]\times\left[\theta(p_0)-\theta(-p_0)\right],
\label{A0S}\\
B_1^{I=0}(p_0)&=&
\left[
b_0\cdot m_s (p_0)^{10}+b_1\cdot \sc(p_0)^8
\right.\cr
&&
+b_2\cdot\sGs(p_0)^6\cr
&&
\left.
+b_3\cdot m_s\qcond^2(p_0)^4+b_4\cdot\sc\GG(p_0)^4
\right]\cr
&&\times\left[\theta(p_0)-\theta(-p_0)\right],
\label{B0S}
\eey
where the coefficients $a_i$ and $b_i$ are given by
\bey
&&a_0=\frac{1}{5^2\cdot 3^2 \cdot 2^{19}\pi^8},\quad a_1=\frac{-7}{5\cdot 3^4\cdot 2^{18}\pi^6},\quad a_2=\frac{1}{3^3\cdot 2^{13}\pi^6},\cr
&&a_3=\frac{-37}{5\cdot3^2\cdot 2^{16}\pi^6},\quad a_4=\frac{1}{5\cdot 3^2\cdot 2^{9} \pi^4},\\
&&b_0=\frac{1}{7\cdot 5^2\cdot 3\cdot 2^{15}\pi^8},\quad b_1=\frac{-1}{3^3\cdot 2^{14}\pi^6},\quad b_2=\frac{101}{5\cdot 3^3\cdot 2^{16}\pi^6},\cr
&&b_3=\frac{1}{3^2\cdot 2^9\pi^4},\quad b_4=\frac{-7}{3^3\cdot 2^{15}\pi^4}.
\eey
In \eqs{A0S}{B0S}, $m_s$ is the strange quark mass, $g$ is the gauge coupling constant and $\alpha_s=g^2/(4\pi)$.
$q\equiv u=d$, $G_{\mu\nu}^a$ is the strength of gluon field and $\bra 0|{\cal O}|0\ket$ denotes
the vacuum expectation value of the operator ${\cal O}$.

The results when we employ ${\eta}_{2,\mu}^{I=0}(x)$ are as follows,
\bey
A_2^{I=0}(p_0)&=&
\left[a_0\cdot (p_0)^{11}+a_1\cdot  \GG(p_0)^7
\right.\cr
&&
+a_2\cdot  m_{s}\sc(p_0)^{7}+a_3\cdot m_{s}\sGs(p_0)^{5}\cr
&&\left.
-9\cdot a_4\cdot \qcond^2 (p_0)^{5}
\right]\times\left[\theta(p_0)-\theta(-p_0)\right],
\label{A0P}\\
B_2^{I=0}(p_0)&=&
\left[
-b_0\cdot m_s (p_0)^{10}-b_1\cdot \sc(p_0)^8
\right.\cr
&&
-b_2\cdot\sGs(p_0)^6\cr
&&
\left.
+7\cdot b_3\cdot m_s\qcond^2(p_0)^4-b_4\cdot\sc\GG(p_0)^4
\right]\cr
&&\times\left[\theta(p_0)-\theta(-p_0)\right],
\label{B0P}
\eey
Let us compare the results for ${\eta}_{1,\mu}^{I=0}$, \eqs{A0S}{B0S}, with those for ${\eta}_{2,\mu}^{I=0}$, \eqs{A0P}{B0P}.
We can see that the OPE convergence of the correlation function of ${\eta}_{2,\mu}^{I=0}$ is clearly slower than that of ${\eta}_{1,\mu}^{I=0}$
since the contributions of dimension 6 ($\qcond^2$) and dimension 7 ($m_s\qcond^2$) terms in \eqs{A0P}{B0P}
are larger than those in \eqs{A0S}{B0S}.
Therefore, we adopt ${\eta}_{1,\mu}^{I=0}$ for deriving the sum rules.

Next, we show the results of OPE for $I=1$ channel.
When using ${\eta}_{1,\mu}^{I=1}(x)$, we obtain
\bey
A_1^{I=1}(p_0)&=&
\left[a_0\cdot(p_0)^{11}+a_1\cdot\GG(p_0)^7
\right.\cr
&&
+a_2\cdot m_{s}\sc(p_0)^{7}
+a_3\cdot m_{s}\sGs(p_0)^{5}\cr
&&\left.
+9\cdot a_4\cdot \qcond^2(p_0)^{5}
\right]\times\left[\theta(p_0)-\theta(-p_0)\right],
\label{A1S}\\
B_1^{I=1}(p_0)&=&
\left[
b_0\cdot m_s (p_0)^{10}+b_1\cdot\sc(p_0)^8
\right.\cr
&&
+b_2\cdot\sGs(p_0)^6\cr
&&\left.
+7\cdot b_3\cdot m_s\qcond^2(p_0)^4
+b_4\cdot\sc\GG(p_0)^4
\right]\cr
&&\times\left[\theta(p_0)-\theta(-p_0)\right],
\label{B1S}
\eey
while the results for ${\eta}_{2,\mu}^{I=1}(x)$ are as follows,
\bey
A_2^{I=1}(p_0)&=&
\left[a_0\cdot(p_0)^{11}+a_1\cdot\GG(p_0)^7
\right.\cr
&&
+a_2\cdot m_{s}\sc(p_0)^{7}
+a_3\cdot m_{s}\sGs(p_0)^{5}\cr
&&\left.
+a_4\cdot \qcond^2(p_0)^{5}
\right]\times\left[\theta(p_0)-\theta(-p_0)\right],
\label{A1P}\\
B_2^{I=1}(p_0)&=&
\left[
-b_0\cdot m_s (p_0)^{10}-b_1\cdot\sc(p_0)^8
\right.\cr
&&
-b_2\cdot\sGs(p_0)^6\cr
&&\left.
-b_3\cdot m_s\qcond^2(p_0)^4
-b_4\cdot\sc\GG(p_0)^4
\right]\cr
&&\times\left[\theta(p_0)-\theta(-p_0)\right].
\label{B1P}
\eey
Comparing \eqs{A1S}{B1S} with \eqs{A1P}{B1P},
we see that ${\eta}_{2,\mu}^{I=1}$ has better convergence than ${\eta}_{1,\mu}^{I=1}$
since the contributions of dimension 6 ($\qcond^2$) and dimension 7 ($m_s\qcond^2$) terms in \eqs{A1S}{B1S}
are larger than those in \eqs{A1P}{B1P}.
From this reason, we employ ${\eta}_{2,\mu}^{I=1}$ to derive the sum rules for $I=1$ states. 

Note here that \eq{A1P} exactly coincides with \eq{A0S}.
Also, \eq{B1P} and \eq{B0S} coincide except for the overall sign.

\subsection{Sum rules for $I=0$ states}
Using the OPE results, \meq{A0S}{B0S}, we derive the QCD sum rules for the $IJ^P=0\frac{3}{2}^\pm$ states of the pentaquark.
From \eq{Spectral-positiveeta}, $\rho_\pm^{\rm OPE}(p_0)$ for $\eta_\mu^{I=0}$ is written in terms of $A(p_0)$ and $B(p_0)$ as,
\bey
\rho_{\pm}^{\rm OPE}(p_0)&=&A^{I=0}(p_0)\pm B^{I=0}(p_0),
\label{OPE0}
\eey
We substitute \eq{OPE0} with \eqs{A0S}{B0S} into the right hand side of \eq{residueSR} to obtain the sum rules,
\bey
&&|\lambda^{I=0}_\pm|^2\exp\left[-{\left(m^{I=0}_\pm\right)^2\over M^2}\right]\cr
&=&
\left[a_0\cdot f(M,\omega^{I=0}_\pm;11)
+a_1\cdot  \GG f(M,\omega^{I=0}_\pm;7)
\right.\cr
&&
+a_2\cdot m_{s}\sc f(M,\omega^{I=0}_\pm;7)+a_3\cdot m_{s}\sGs f(M,\omega^{I=0}_\pm;5)\cr
&&\left.
+a_4\cdot \qcond^2 f(M,\omega^{I=0}_\pm;5)
\right]\cr
&&
\pm\left[
b_0\cdot m_s f(M,\omega^{I=0}_\pm;10)+b_1\cdot\sc f(M,\omega^{I=0}_\pm;8)
\right.\cr
&&
+b_2\cdot\sGs f(M,\omega^{I=0}_\pm;6)+b_3\cdot m_s\qcond^2 f(M,\omega^{I=0}_\pm;4)\cr
&&\left.
+b_4\cdot \sc\GG f(M,\omega^{I=0}_\pm;4)
\right],
\label{res0}
\eey
where $|\lambda^{I}_\pm|^2$, $m^{I}_\pm$ and $\omega^{I}_\pm$ are the pole residues, the masses and the effective continuum threshold with the isospin $I$ channel, respectively. 
$f(M,\omega;n)$ is the integral defined by 
\bey
f(M,\omega;n)\equiv\int_{0}^{\omega}dp_0 (p_0)^n \exp(-{p_0^2\over M^2}).
\eey

The sum rules for the masses are obtained by substituting the OPE into \eq{massSR} as
\bey
(m_\pm^{I=0})^2&=&
\left\{
\left[a_0\cdot f(M,\omega^{I=0}_\pm;13)
+a_1\cdot\GG f(M,\omega^{I=0}_\pm;9)\right.\right.\cr
&&
+a_2\cdot m_{s}\sc f(M,\omega^{I=0}_\pm;9)+a_3\cdot m_{s}\sGs f(M,\omega^{I=0}_\pm;7)\cr
&&\left.
+a_4\cdot\qcond^2 f(M,\omega^{I=0}_\pm;7)
\right]\cr
&&
\pm\left[
b_0\cdot m_s f(M,\omega^{I=0}_\pm;12)
+b_1\cdot \sc f(M,\omega^{I=0}_\pm;10)
\right.\cr
&&
+b_2\cdot \sGs f(M,\omega^{I=0}_\pm;8)+b_3\cdot m_s\qcond^2 f(M,\omega^{I=0}_\pm;6)\cr
&&\left.\left.
+b_4\cdot \sc\GG f(M,\omega^{I=0}_\pm;6)
\right]\right\}\cr
&\Big/&\left\{
\left[a_0\cdot f(M,\omega^{I=0}_\pm;11)
+a_1\cdot \GG f(M,\omega^{I=0}_\pm;7)
\right.\right.\cr
&&
+a_2\cdot m_{s}\sc f(M,\omega^{I=0}_\pm;7)+a_3\cdot m_{s}\sGs f(M,\omega^{I=0}_\pm;5)\cr
&&\left.
+a_4\cdot\qcond^2 f(M,\omega^{I=0}_\pm;5)
\right]\cr
&&
\pm\left[
b_0\cdot m_s f(M,\omega^{I=0}_\pm;10)
+b_1\cdot\sc f(M,\omega^{I=0}_\pm;8)
\right.\cr
&&
+b_2\cdot \sGs f(M,\omega^{I=0}_\pm;6)+b_3\cdot m_s\qcond^2 f(M,\omega^{I=0}_\pm;4)\cr
&&\left.\left.
+b_4\cdot\sc\GG f(M,\omega^{I=0}_\pm;4)
\right]\right\}.
\label{mass0}
\eey
It should be noted that the mass splitting between positive and negative parity states is due to the $\pm$ sign of the terms containing $m_s$ and the condensates of chiral odd operators \cite{jido}.
\subsection{Sum rules for $I=1$ states}
From \eq{Spectral-negativeeta}, we see that $\rho_\pm^{\rm OPE}(p_0)$ for $\eta_\mu^{I=1}$ is written in terms of $A(p_0)$ and $B(p_0)$ as,
\bey
\rho_{\pm}^{\rm OPE}(p_0)&=&A^{I=1}(p_0)\mp B^{I=1}(p_0).
\label{OPE1}
\eey
The sum rules for the $IJ^P=1\frac{3}{2}^\pm$ states are obtained by using \eq{OPE1} with \eqs{A1P}{B1P},
\bey
&&|\lambda^{I=1}_\pm|^2\exp\left[-{\left(m^{I=1}_\pm\right)^2\over M^2}\right]\cr
&=&\left[a_0\cdot f(M,\omega^{I=1}_\pm;11)
+a_1\cdot \GG f(M,\omega^{I=1}_\pm;7)
\right.\cr
&&
+a_2\cdot m_{s}\sc f(M,\omega^{I=1}_\pm;7)+a_3\cdot m_{s}\sGs f(M,\omega^{I=1}_\pm;5)\cr
&&\left.
+a_4\cdot\qcond^2 f(M,\omega^{I=1}_\pm;5)
\right]\cr
&&\mp
\left[-b_0\cdot m_s f(M,\omega^{I=1}_\pm;10)-b_1\cdot \sc f(M,\omega^{I=1}_\pm;8)
\right.\cr
&&
-b_2\cdot \sGs f(M,\omega^{I=1}_\pm;6)-b_3\cdot m_s\qcond^2 f(M,\omega^{I=1}_\pm;4)\cr
&&\left.
-b_4\cdot\sc\GG f(M,\omega^{I=1}_\pm;4)
\right].
\label{res1}
\eey
The mass sum rule for $I=1$ are as follows,
\bey
(m_\pm^{I=1})^2&=&
\left\{
\left[a_0\cdot f(M,\omega^{I=0}_\pm;13)
+a_1\cdot\GG f(M,\omega^{I=0}_\pm;9)\right.\right.\cr
&&
+a_2\cdot m_{s}\sc f(M,\omega^{I=0}_\pm;9)+a_3\cdot m_{s}\sGs f(M,\omega^{I=0}_\pm;7)\cr
&&\left.
+a_4\cdot\qcond^2 f(M,\omega^{I=0}_\pm;7)
\right]\cr
&&
\mp\left[
-b_0\cdot m_s f(M,\omega^{I=0}_\pm;12)
-b_1\cdot \sc f(M,\omega^{I=0}_\pm;10)
\right.\cr
&&
-b_2\cdot \sGs f(M,\omega^{I=0}_\pm;8)-b_3\cdot m_s\qcond^2 f(M,\omega^{I=0}_\pm;6)\cr
&&\left.\left.
-b_4\cdot \sc\GG f(M,\omega^{I=0}_\pm;6)
\right]\right\}\cr
&\Big/&\left\{
\left[a_0\cdot f(M,\omega^{I=0}_\pm;11)
+a_1\cdot \GG f(M,\omega^{I=0}_\pm;7)
\right.\right.\cr
&&
+a_2\cdot m_{s}\sc f(M,\omega^{I=0}_\pm;7)+a_3\cdot m_{s}\sGs f(M,\omega^{I=0}_\pm;5)\cr
&&\left.
+a_4\cdot\qcond^2 f(M,\omega^{I=0}_\pm;5)
\right]\cr
&&
\mp\left[
-b_0\cdot m_s f(M,\omega^{I=0}_\pm;10)
-b_1\cdot\sc f(M,\omega^{I=0}_\pm;8)
\right.\cr
&&
-b_2\cdot \sGs f(M,\omega^{I=0}_\pm;6)-b_3\cdot m_s\qcond^2 f(M,\omega^{I=0}_\pm;4)\cr
&&\left.\left.
-b_4\cdot\sc\GG f(M,\omega^{I=0}_\pm;4)
\right]\right\}.
\label{mass1}
\eey  

\section{Numerical Results}
In this section, we show the numerical results obtained from the sum rules for $IJ^P=0\frac{3}{2}^\pm$ states, \eqs{res0}{mass0}, and those for $1\frac{3}{2}^\pm$, \eqs{res1}{mass1}.
Throughout this paper, we use the QCD parameters of the standard values, 
$\qcond=(-0.23\;{\rm GeV})^3$,
$m_s=0.12\;{\rm GeV}$, $\sc=0.8\qcond$,
$\sGs=(0.8\;{\rm GeV}^2)\sc$, $\GG=(0.33\;{\rm GeV})^4$.

\subsection{$I=0$ states}
\begin{figure}
{\includegraphics[width=13cm,keepaspectratio]{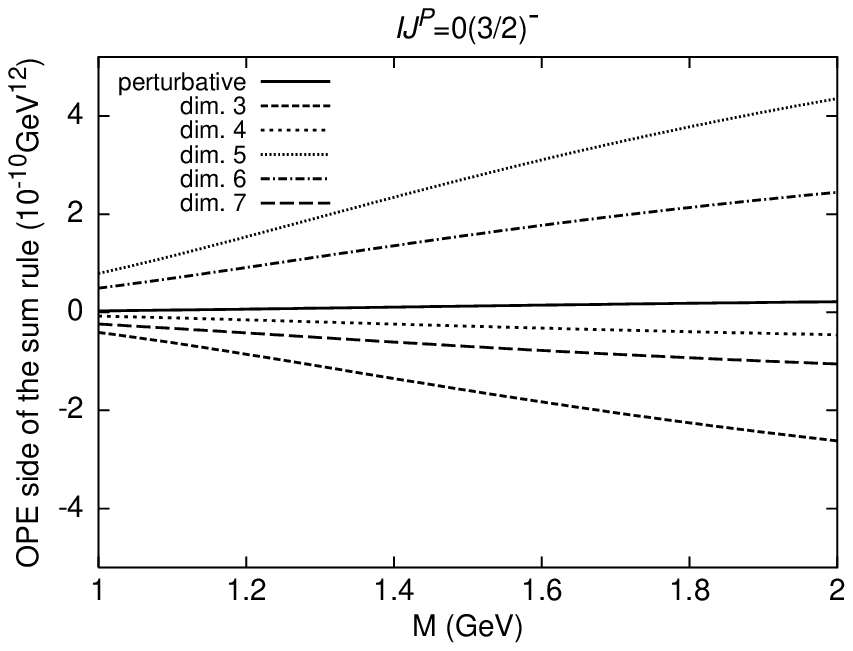}}
\caption{Contributions of the terms in OPE side (the right hand side) of \eq{res0} for negative parity as functions of $M$ with the effective continuum threshold $\omega^{I=0}_-=1.8\,{\rm GeV}$.}
\label{OPE0n}
\end{figure}

\begin{figure}
{\includegraphics[width=13cm,keepaspectratio]{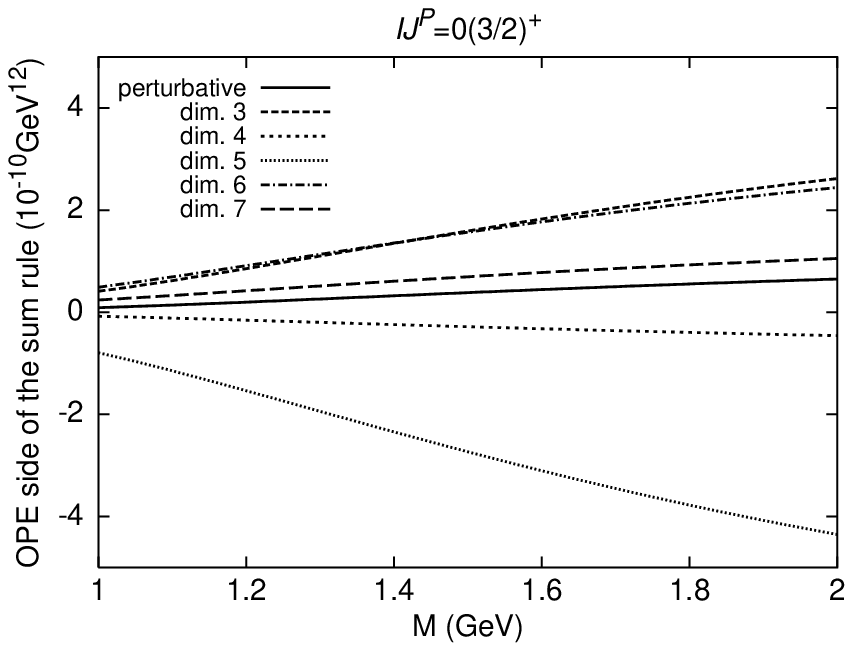}}
\caption{Contributions of the terms in OPE side (the right hand side) of \eq{res0} for positive parity as functions of $M$ with the effective continuum threshold $\omega^{I=0}_+=1.8\,{\rm GeV}$.}
\label{OPE0p}
\end{figure}

First, in order to see how good the convergence of OPE is, we show the contribution of each term in the right hand sides of the sum rules, \eq{res0}, in Figs.\ref{OPE0n} and \ref{OPE0p}.
We see that dimension 5 term, namely, mixed condensate $\sGs$ gives dominant contribution.
The terms higher than dimension 5 seems to decrease with increasing dimension; 
the contributions of dimension 6 and 7 terms are about 50\% and 30\%
of dimension 5 terms, respectively. 

\begin{figure}
{\includegraphics[width=13cm,keepaspectratio]{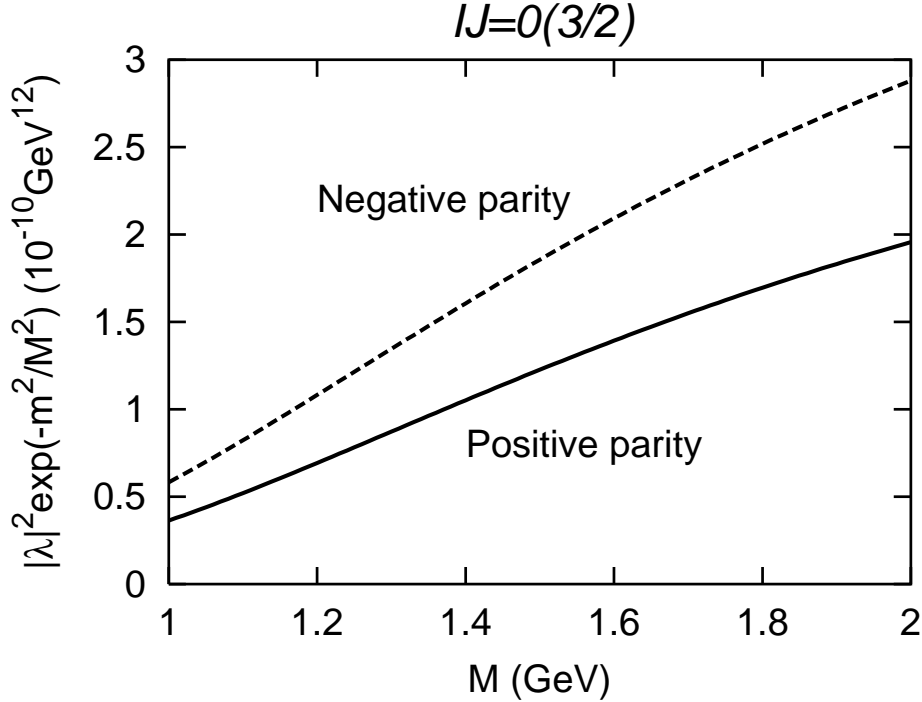}}
\caption{$|\lambda_\mp|^2\exp(-{{m_\mp}^2\over M^2})$ for $I=0$ as functions of Borel mass, $M$, obtained from \eq{res0} with the effective continuum threshold $\omega^{I=0}_\mp=1.8\,{\rm GeV}$.}
\label{res0-M}
\end{figure}

We plotted in Fig.\ref{res0-M} the right-hand side of \eq{res0} as functions of the Borel mass, $M$.
The pole residue $|\lambda_{\mp}^{I=0}|^2$ must be positive
otherwise the pole of the pentaquark is spurious.
As can be seen in Fig.\ref{res0-M}, the right-hand side of \eq{res0} is positive. This suggests that the present QCD sum rule does not exclude
the exisitence of the $0\frac{3}{2}^{\pm}$ pentaquarks.

\begin{figure}
{\includegraphics[width=13cm,keepaspectratio]{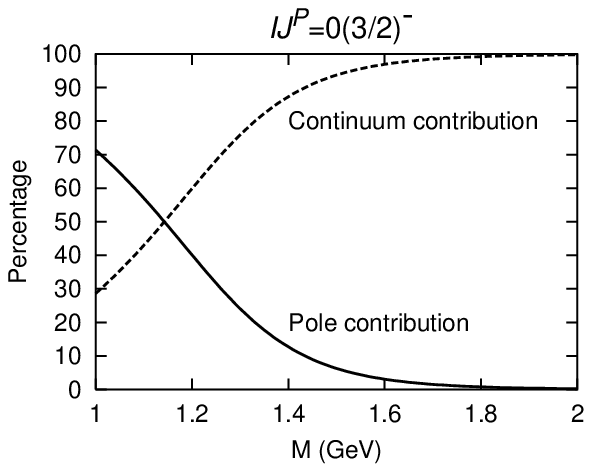}}
\caption{Relative magnitude of the pole (solid line) and the continuum (dashed line) contribution in \eq{res0} for negative parity as functions of $M$ with the effective continuum threshold $\omega^{I=0}_-=1.8\,{\rm GeV}$.}
\label{polecont0n}
\end{figure}

\begin{figure}
{\includegraphics[width=13cm,keepaspectratio]{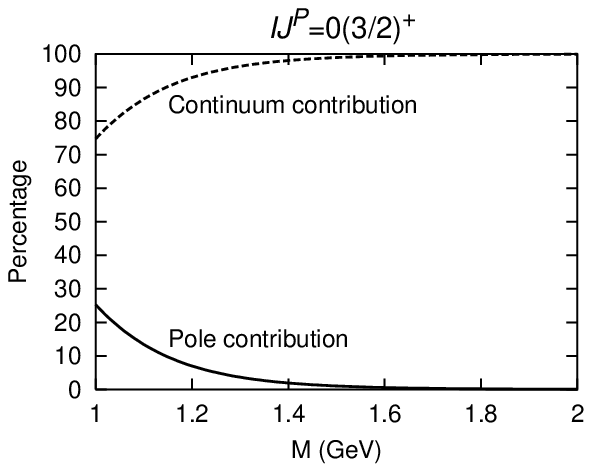}}
\caption{Relative magnitude of the pole (solid line) and the continuum (dashed line) contribution in \eq{res0} for positive parity as functions of $M$ with the effective continuum threshold $\omega^{I=0}_+=1.8\,{\rm GeV}$.}
\label{polecont0p}
\end{figure}

In Figs.\ref{polecont0n} and \ref{polecont0p}, we show the relative strength of the pole and the continuum contribution for negative and positive parity, respectively.
The pole contribution is given by \eq{res0},
while the continuum contribution is obtained by subtracting \eq{res0} from \eq{res0} with $\omega^{I=0}_\mp=\infty$.
The percentage of the pole contribution must be as large as possible, but one usually accepts values around $50\%$.
In the sum rule for negative parity, the pole contribution is sufficiently large
at around $M=1.2\,{\rm GeV}$. On the other hand, in the sum rule for positive parity, the continuum contribution dominates the pole contribution, 
which implies that the sum rule for positive parity is less reliable than that for negative parity.   

\begin{figure}
{\includegraphics[width=13cm,keepaspectratio]{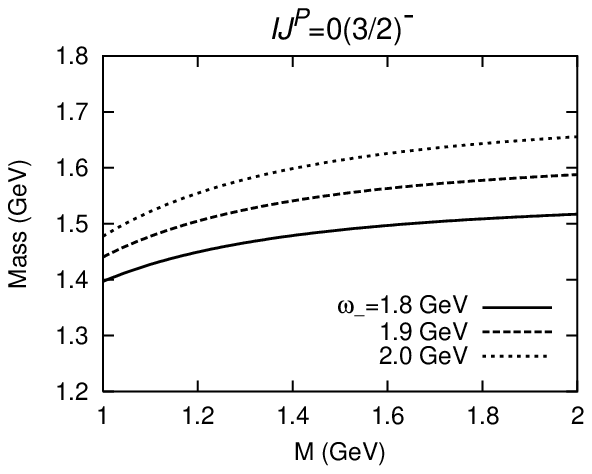}}
\caption{Mass of the $IJ^P=0\frac{3}{2}^-$ pentaquark as a function of $M$ with the effective continuum threshold $\omega^{I=0}_{-}=1.8\,{\rm GeV}$ (solid line), $1.9\,{\rm GeV}$ (long-dashed), $2.0\,{\rm GeV}$ (short-dashed).}
\label{mass0-}
\end{figure}

In Fig.\ref{mass0-}, we plotted the mass of the $0\frac{3}{2}^-$ state, $m_-^{I=0}$, against the Borel mass which is obtained from \eq{mass0}.
We see that there exists a region where the dependence on the Borel mass is weak and therefore the sum rule works.
However, the curve depends on the choice of the effective continuum thershold, $\omega^{I=0}_-$.
The continuum which comes from $NK$ states
starts up very gradually since it is $D$-wave in this channel.
Around $1.8$ GeV, the $S$-wave $N K^*$ state
opens and above the threshold it may give a large contribution to the continuum.
We therefore choose the values of the effective continuum thershold as $\omega^{I=0}_{-}=1.8$, $1.9$, $2.0\,{\rm GeV}$.
From the region of the curve weakly dependent of $M$,
$m_-^{I=0}$ is predicted to be $1.5\sim 1.7$ GeV.
A remarkable point here is that 
the mass is below the $N K^*$ threshold ($1.83$ GeV).
This means that the $0\frac{3}{2}^-$ pentaquark can be extremely narrow, as discussed in the next section.

\begin{figure}
{\includegraphics[width=13cm,keepaspectratio]{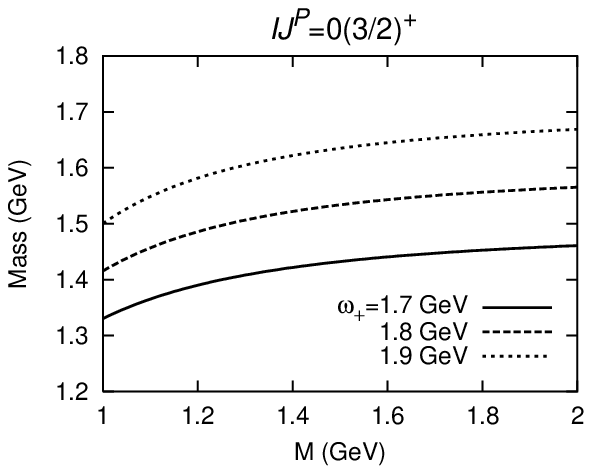}}
\caption{Mass of $IJ^P=0\frac{3}{2}^+$ pentaquark as a function of $M$ with the effective continuum threshold $\omega^{I=0}_{+}=1.7\,{\rm GeV}$ (solid line), $1.8\,{\rm GeV}$ (long-dashed), $1.9\,{\rm GeV}$ (short-dashed).}
\label{mass0+}
\end{figure}

The mass of the $0\frac{3}{2}^+$ state, $m_+^{I=0}$, against the Borel mass is shown in Fig.\ref{mass0+}.
The continuum in this channel mainly comes from the $P$-wave $NK$ states. It starts up more gradually than the $S$-wave $NK$ continuum,
but more rapidly than the $D$-wave continuum.
In view of this point, we choose $\omega^{I=0}_{+}=1.7 $, $1.8$, $1.9\,{\rm GeV}$.
We see from Figs.\ref{mass0-} and \ref{mass0+} that
the dependence of the curve for $0\frac{3}{2}^+$ on the effective continuum threshold is stronger than that for $0\frac{3}{2}^-$.
This is natural since the relative strength of the pole for $0\frac{3}{2}^+$ is weak compared with that for $0\frac{3}{2}^-$, as can be seen in Figs.\ref{polecont0n} and \ref{polecont0p}.

The masses of positive and negative parity states are nearly degenerate
for $0\frac{3}{2}$ pentaquark.
In the QCD sum rules for baryons, the parity splitting  originates in the chiral odd term.
For the $0\frac{3}{2}$ pentaquark, the contribution of the chiral odd terms in the sum rule, \eq{res0}, are small compared with the chiral even terms, which leads to the degeneracy of the positive and negative parity states.


\subsection{$I=1$ states}
\begin{figure}
{\includegraphics[width=13cm,keepaspectratio]{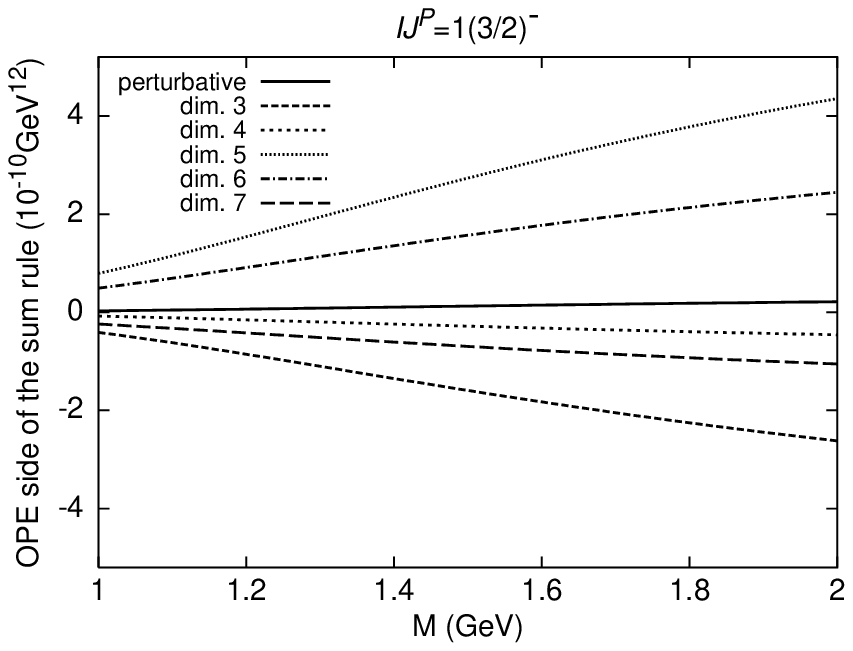}}
\caption{Contributions of the terms in OPE side (the right hand side) of \eq{res1} for negative parity as functions of $M$ with the effective continuum threshold $\omega^{I=1}_-=1.8\,{\rm GeV}$.}
\label{OPE1n}
\end{figure}

\begin{figure}
{\includegraphics[width=13cm,keepaspectratio]{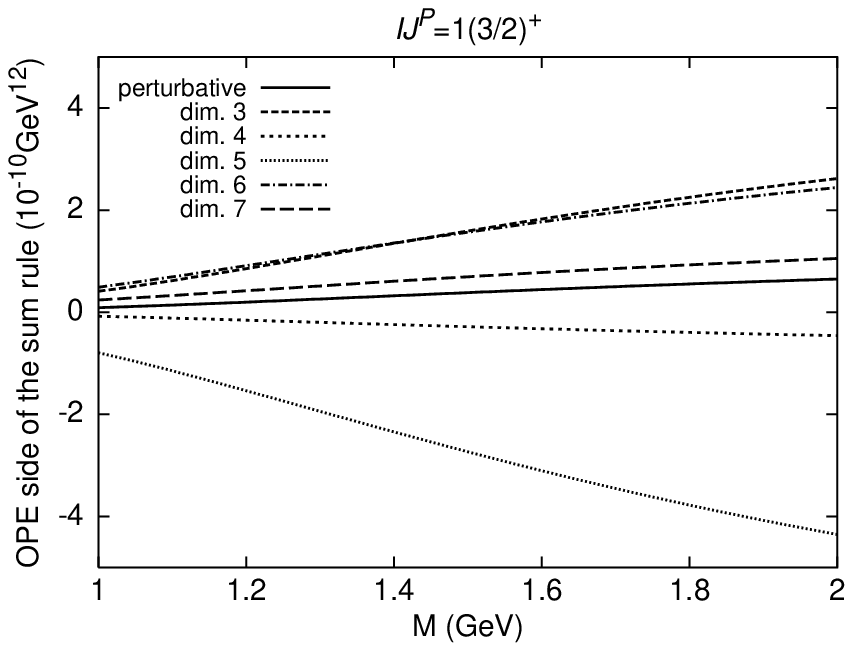}}
\caption{Contributions of the terms in OPE side (the right hand side) of \eq{res1} for positive parity as functions of $M$ with the effective continuum threshold $\omega^{I=1}_+=1.8\,{\rm GeV}$.}
\label{OPE1p}
\end{figure}

We plot the contribution of each OPE term in the right hand sides of the sum rules, \eq{res1}, in Figs.\ref{OPE1n} and \ref{OPE1p}.
The behavior of the OPE for this channel is the same as that for $I=0$ channel.

\begin{figure}
{\includegraphics[width=13cm,keepaspectratio]{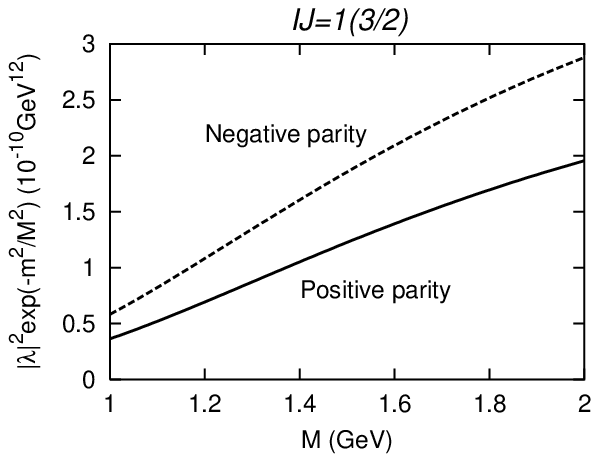}}
\caption{$|\lambda_\pm|^2\exp(-{{m_\pm}^2\over M^2})$ for $I=1$ as functions of $M$ obtained from \eq{res0} with the effective continuum threshold $\omega^{I=1}_\pm=1.8\,{\rm GeV}$.}
\label{res1-M}
\end{figure}

The right-hand side of \eq{res1} as functions of $M$ are plotted in Fig.\ref{res1-M}, which shows that the pole residues of the $I=1$ states
are positive.
This suggests that the poles of the $1\frac{3}{2}^{\pm}$ pentaquark
are not spurious.

\begin{figure}
{\includegraphics[width=13cm,keepaspectratio]{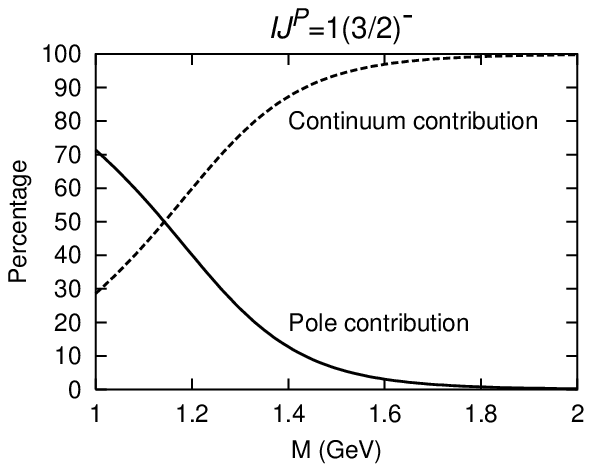}}
\caption{Relative magnitude of the pole (solid line) and the continuum (dashed line) contribution in \eq{res1} for negative parity as functions of $M$ with the effective continuum threshold $\omega^{I=1}_-=1.8\,{\rm GeV}$.}
\label{polecont1n}
\end{figure}

\begin{figure}
{\includegraphics[width=13cm,keepaspectratio]{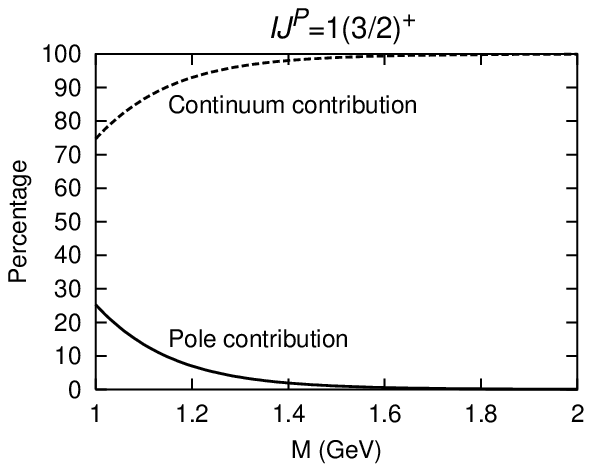}}
\caption{Relative magnitude of the pole (solid line) and the continuum (dashed line) contribution in \eq{res1} for positive parity as functions of $M$ with the effective continuum threshold $\omega^{I=1}_+=1.8\,{\rm GeV}$.}
\label{polecont1p}
\end{figure}

In Figs.\ref{polecont1n} and \ref{polecont1p}, we show the relative strength of the pole and the continuum contribution for negative and positive parity, respectively.
The pole contribution in the sum rule for negative parity is sufficiently large
at around $M=1.2\,{\rm GeV}$, while for positive parity the continuum contribution dominates over the pole contribution, 
which implies that the reliability of the sum rule for positive parity is lower than that for negative parity.   


\begin{figure}
{\includegraphics[width=13cm,keepaspectratio]{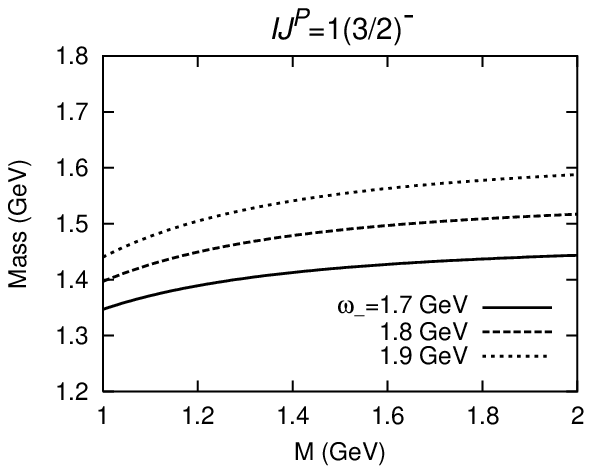}}
\caption{Mass of $IJ^P=1\frac{3}{2}^-$ pentaquark as a function of $M$ with the effective continuum threshold $\omega^{I=1}_{-}=1.7\,{\rm GeV}$ (solid line), $1.8\,{\rm GeV}$ (long-dashed), $1.9\,{\rm GeV}$ (short-dashed).}
\label{mass1-}
\end{figure}

In Fig.\ref{mass1-}, we plotted the mass of the $1\frac{3}{2}^-$ state, 
$m_-^{I=1}$, against the Borel mass which is obtained from \eq{mass1}.
Although the dependence on the Borel mass is weak,
the curve depends on the choice of the effective continuum thershold $\omega^{I=1}_-$.
The continuum which comes from $D$-wave $NK$ states
starts up very gradually.
Around $1.7$ GeV, the $S$-wave $\Delta K$ state opens and above the threshold the continuum of the $S$-wave $\Delta K$ may give a large contribution.
Thus we choose $\omega^{I=1}_{-}=1.7$, $1.8$, $1.9\,{\rm GeV}$.
From the stabilized region of the curve,
we predict the mass to be $ 1.4\sim 1.6$ GeV,
which is below the $\Delta K$ threshold ($1.73$ GeV).

\begin{figure}
{\includegraphics[width=13cm,keepaspectratio]{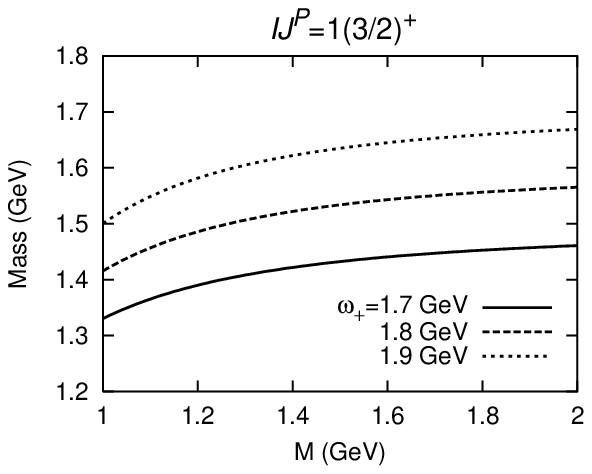}}
\caption{Mass of $IJ^P=1\frac{3}{2}^+$ pentaquark as a function of $M$ with the effective continuum threshold $\omega^{I=1}_{+}=1.7\,{\rm GeV}$ (solid line), $1.8\,{\rm GeV}$ (long-dashed), $1.9\,{\rm GeV}$ (short-dashed).}
\label{mass1+}
\end{figure}

Let us turn to the sum rule for the positive parity state.
The mass against the Borel mass is shown
in Fig.\ref{mass1+}.
The continuum in this channel mainly comes from the $P$-wave $NK$ states.
The effective continuum threshold is taken to be the same value as that for 
the $0\frac{3}{2}^+$ state: $\omega^{I=1}_{+}=1.7 $, $1.8$, $1.9\,{\rm GeV}$.

As in the case for the $IJ=0\frac{3}{2}$ pentaquark, positive and negative parity states are nearly degenerate for $IJ=1\frac{3}{2}$.
In the sum rules for the $1\frac{3}{2}$ pentaquark, \eq{res1}, the contribution of  chiral odd terms are small compared with the chiral even terms, which leads to the degeneracy of the positive and negative parity states.

\section{discussion}


\subsection{$J^P=\frac{3}{2}^-$ states} 
We have found that positive and negative parity states of the pentaquark baryons with high spin may exist in low mass region,
which is not the case for ordinary three-quark baryons.
The widths of these states are sensitive to the mass positions relative to
the threshold of $S$-wave meson-baryon decay.
For the $0\frac{3}{2}^-$ and the $1\frac{3}{2}^-$ states, no $S$-wave channel of meson-baryon decay is open, if their masses are below $NK^*$ and $\Delta K$ threshold, respectively.
In the present analysis, we obtained $1.5\sim 1.7$ GeV for the mass of the $0\frac{3}{2}^-$ state, which is below the $NK^*$ threshold energy.
On the other hand, the predicted mass of the $1\frac{3}{2}^-$ state, $1.4\sim 1.6$ GeV, is also below the $\Delta K$ threshold.
This means that both states can decay only to the $D$-wave $NK$ states because of the conservation law of total spin and parity.
Due to the high centrifugal barrier,  the widths are strongly suppressed.   
As a result, these states may be observed as narrow peaks.
The $0\frac{3}{2}^-$ state can be a candidate of the observed $\Theta^+(1540)$ and the $1\frac{3}{2}^-$ pentaquark might be a new particle, $\Theta^{++}$.

The possibility of high spin states of the pentaquark has also been suggested from other approaches.
Very recently, by employing a quark model with the meson exchange and one-gluon exchange interaction, negative parity $uudd\bar s$ pentaquarks have been investigated \cite{takeuchi}.
The low lying states found in this calculation are
$0\frac{1}{2}^-$ and $0\frac{3}{2}^-$ states.
The former may be broad and not be observed since it lies above the $NK$ threshold.
On the other hand, the latter is below the $NK^*$ threshold and therefore
it can be observed as a narrow peak.
 
The existence of $1\frac{3}{2}^-$ pentaquark as a low lying state
has been suggested in Ref.\cite{amdpenta}.
In Ref.\cite{amdpenta}, a simple quark model
in which constituent quarks interact via one-gluon exchange force at short distances and confining (or string) potential at long distances was considered.
A $qqqq\bar{q}$ system has a connected string configuration corresponding to a confined state, in addition to an ordinary meson-baryon like configuration.
A variational method called AMD (antisymmetrized
molecular dynamics) \cite{amd1,amd2} was applied to the confined five-body system with $uudd\bar{s}$ 
and all the possible spin parity states with the connected string configurations were calculated. 
The narrow and low lying states they have found are $0\frac{1}{2}^+$,
$0\frac{3}{2}^+$ and $1\frac{3}{2}^-$ states.
The former two states have just the same structure as that conjectured by Jaffe and Wilczek \cite{jaffe}. 
We represent it as $[ud]_{S=0,I=0}[ud]_{S=0,I=0}[\bar{s}]$,
where $[ud]_{S,I}$ denotes a color $\bf\bar 3$ $ud$-diquark with spin $S$ and isospin $I$. 
Both of the two diquarks gain color magnetic interaction since they have $S=0$. However, the $0\frac{1}{2}^+$ and 
$0\frac{3}{2}^+$ states lose kinetic and string energy
since the two diquarks, which are to be antisymmetric in color, are identical and since they must be relatively $P$-wave.
In Ref.\cite{amdpenta}, another energetically favorable state
has been predicted,
which consists of an $S=0$ diquark and an $S=1$ diquark:
$[ud]_{S=0,I=0}[ud]_{S=1,I=1}[\bar{s}]$.
The quantum number of this state is totally $1\frac{3}{2}^-$.
It loses color magnetic interaction due to the existence of a diquark with $S=1$.
However, the $1\frac{3}{2}^-$ state gains kinetic and string energy since the two diquarks are no longer identical and since they can be relatively $S$-wave.
Owing to the balance between the energy gain and the loss, the $1\frac{3}{2}^-$ state degenerate with the $0\frac{1}{2}^+$ and $0\frac{3}{2}^+$ states.

A problem common to the above two works
is that within the quark models one cannot predict the absolute masses but only the level structure of the pentaquarks in principle.
The quark models employed in Refs.\cite{takeuchi,amdpenta} relies on the 
zero-point energy of the confining potential.
The value of the zero-point energy, however, is not determined within this kind of emprical models.
In Ref.\cite{amdpenta}, it was adjusted to reproduce the observed mass of $\Theta^+$.
Whereas, the QCD sum rule is able to estimate the absolute mass
though it depends on the effective continuum threshold.
We confirmed from the QCD sum rule that the $0\frac{3}{2}^-$ and the $1\frac{3}{2}^-$ states are below the $S$-wave threshold and therefore
they can be narrow states.

The pentaquark with $1\frac{3}{2}^-$ has also been found to exist as a resonant state in the $\Delta K$ channel \cite{oset} from the chiral unitary approach.
This state is generated due to an attractive interaction in that channel existing in the lowest order chiral Lagrangian.
The attractive interaction leads to a pole of the complex energy plane and manifests itself in a large strength of the $\Delta K$ scattering amplitude with $L=0$ and $I=1$.
We note that the interpolating field, \eq{currentisovectorP}, can also couple 
with such a $\Delta K$ resonance states 
because it contains the $\Delta K$ component as is shown by Fierz transformation.

\subsection{$J^P=\frac{3}{2}^+$ states} 
The $0\frac{3}{2}^+$ state has been discussed as a $LS$ partner of 
the $0\frac{1}{2}^+$ state \cite{close}. 
In the QCD sum rule study of the $0\frac{1}{2}^+$ state \cite{eide},
it was found in the energy region compatible with the experimentally measured $\Theta^+$ mass.
In Ref.\cite{eide}, the interpolating field based on Jaffe and Wilczek's conjecture:
$[ud]_{S=0,I=0}[ud]_{S=0,I=0}[\bar{s}]$ for the $0\frac{1}{2}^+$ state was employed.
If such a diquark structure is realized,
the mass splitting between the $0\frac{1}{2}^+$ and the $0\frac{3}{2}^+$ is expected to be small because the effect of the spin-orbit force should be small due to the existence of two spinless diquarks.
In Ref.\cite{close}, the authors predicted that the $0\frac{3}{2}^+$ state 
may exist in the mass region 1.54GeV $\sim1.68$ GeV based on the diquark picture. In the present calculation, since the interporating field for the $I=0$  
can couple with such the diquark configuration,  
the obtained result for the $0\frac{3}{2}^+$ state may be associated with 
the $LS$ partner of the $0\frac{1}{2}^+$ state with 
Jaffe and Wilczek's diquark structure. 
Our result implies a possiblity of the $0\frac{3}{2}^+$
state. 
However, as was shown in the previous section, the sum rule for $0\frac{3}{2}^+$ channel is less reliable.
Therefore we should not discuss the mass difference with the $0\frac{1}{2}^+$
state using the results from the present sum rule.
To do that, it would be necessary to construct the sum rule in which the background continuum contribution is made as small as possible.


The $0\frac{3}{2}^+$ and $1\frac{3}{2}^+$ pentaquarks
are expected to be broader than the $0\frac{3}{2}^-$ and $1\frac{3}{2}^-$ states.
The reason is as follows. The $0\frac{3}{2}^+$ and $1\frac{3}{2}^+$ states can decay into $P$-wave $NK$ states, 
while the $0\frac{3}{2}^-$ and $1\frac{3}{2}^-$ states decay only to $D$-wave $NK$ states. 
The centrifugal barrier of $P$-wave $NK$ states is lower than that of $D$-wave $NK$ states, which makes the positive parity states broader than the negative parity states.
The present result for the $1\frac{3}{2}^+$ state is consistent with a recent calculation by Skyrme model \cite{borisyuk}.
The authors in Ref.\cite{borisyuk} predicted that there exists a new isotriplet
of $\Theta$-baryons with $1\frac{3}{2}^+$.
Its mass is $1595\,{\rm MeV}$ and the width is large: $\Gamma\sim 80\,{\rm MeV}$.

\section{Summary}
In summary, we have studied the high spin ($J=3/2$ ) states of pentaquark
with $I=0$ and $I=1$ using the method of QCD sum rule.
We have derived the QCD sum rules for both of the negative and positive parity states.
The QCD sum rule suggests the existence of pentaquark states with narrow width,  $IJ^P=0\frac{3}{2}^-$ and $1\frac{3}{2}^-$.
The masses for the $I=0$ and $I=1$ states
are predicted to be $1.5\sim 1.7$ GeV and $1.4\sim 1.6$ GeV,
which are much below the $N K^*$ threshold and the 
$\Delta K$ threshold, respectively.
Since only the $D$-wave decay to $NK$ channel is allowed,
they should be extremely narrow states.
Concerning the mass difference between $IJ^P=0\frac{3}{2}^-$ and $1\frac{3}{2}^-$ states, we cannot say anything definitely, because the masses depend on the values of the effective continumm threshold in the present calculation.
The QCD sum rule also shows the possibility of the existence of 
the $J^P=3/2^+$ states.
The positive parity states may be broader than the negative parity states
since they are allowed to decay into $P$-wave $NK$ state.

It is worth mentioning that this is the first QCD sum rule analysis of 
high spin ($J=3/2$) states of the pentaquark. 
The important point is that this work suggests possible existence of 
the high spin states in the same energy region as the $J=1/2$ states 
obtained by QCD sum rule in Ref.\cite{eide}. 
This abnormal spectra of the pentaquark are contrast to the ordinary baryon 
spectra. It is also remarkable that
the exotic spin and parity, $\frac{3}{2}^-$, leads to the existence of extremely narrow states.
It would be interesting to see if lattice calculation could confirm our
findings since most of the existing works using QCD sum rules or lattice QCD 
have concentrated on $J=\frac{1}{2}$ pentaquark states.

Finally, we would like to give a comment on the present formalism of QCD sum rules, which has been widely used for pentaquark.
The formalism is a simple extension of that for the ordinary hadrons.
One of the subtle problems in QCD sum rules for pentaquark is 
how to properly extract a resonance in the contamination of the background continuum states. 
Before we obtain a final conclusion for the pentaquark study with 
the QCD sum rule, it is a necessary process to examine the validity of the QCD sum rule formalism for exotic hadrons.
In order to settle this problem, further experimental and theoretical
studies on excited states and other pentaquark states with their spin and parity are desired. It is also useful to compare the lattice QCD calculations
with the QCD sum rule results. 
\acknowledgements{The authors would like to thank Prof. M. Oka and Dr. T. Doi and J. Sugiyama for letting us to find errors in the OPE, though the conclusion is unchanged.
One of the authors (T. N.) thanks Prof. M. Oka and Dr. N. Ishii for usefull comments and discussions.}


\def\Ref#1{[\ref{#1}]}
\def\Refs#1#2{[\ref{#1},\ref{#2}]}
\def\npb#1#2#3{{Nucl. Phys.\,}{\bf B{#1}},\,#2\,(#3)}
\def\npa#1#2#3{{Nucl. Phys.\,}{\bf A{#1}},\,#2\,(#3)}
\def\np#1#2#3{{Nucl. Phys.\,}{\bf{#1}},\,#2\,(#3)}
\def\plb#1#2#3{{Phys. Lett.\,}{\bf B{#1}},\,#2\,(#3)}
\def\prl#1#2#3{{Phys. Rev. Lett.\,}{\bf{#1}},\,#2\,(#3)}
\def\prd#1#2#3{{Phys. Rev.\,}{\bf D{#1}},\,#2\,(#3)}
\def\prc#1#2#3{{Phys. Rev.\,}{\bf C{#1}},\,#2\,(#3)}
\def\prb#1#2#3{{Phys. Rev.\,}{\bf B{#1}},\,#2\,(#3)}
\def\pr#1#2#3{{Phys. Rev.\,}{\bf{#1}},\,#2\,(#3)}
\def\ap#1#2#3{{Ann. Phys.\,}{\bf{#1}},\,#2\,(#3)}
\def\prep#1#2#3{{Phys. Reports\,}{\bf{#1}},\,#2\,(#3)}
\def\rmp#1#2#3{{Rev. Mod. Phys.\,}{\bf{#1}},\,#2\,(#3)}
\def\cmp#1#2#3{{Comm. Math. Phys.\,}{\bf{#1}},\,#2\,(#3)}
\def\ptp#1#2#3{{Prog. Theor. Phys.\,}{\bf{#1}},\,#2\,(#3)}
\def\ib#1#2#3{{\it ibid.\,}{\bf{#1}},\,#2\,(#3)}
\def\zsc#1#2#3{{Z. Phys. \,}{\bf C{#1}},\,#2\,(#3)}
\def\zsa#1#2#3{{Z. Phys. \,}{\bf A{#1}},\,#2\,(#3)}
\def\intj#1#2#3{{Int. J. Mod. Phys.\,}{\bf A{#1}},\,#2\,(#3)}
\def\sjnp#1#2#3{{Sov. J. Nucl. Phys.\,}{\bf #1},\,#2\,(#3)}
\def\pan#1#2#3{{Phys. Atom. Nucl.\,}{\bf #1},\,#2\,(#3)}
\def\app#1#2#3{{Acta. Phys. Pol.\,}{\bf #1},\,#2\,(#3)}
\def\jmp#1#2#3{{J. Math. Phys.\,}{\bf {#1}},\,#2\,(#3)}
\def\cp#1#2#3{{Coll. Phen.\,}{\bf {#1}},\,#2\,(#3)}
\def\epjc#1#2#3{{Eur. Phys. J.\,}{\bf C{#1}},\,#2\,(#3)}
\def\mpla#1#2#3{{Mod. Phys. Lett.\,}{\bf A{#1}},\,#2\,(#3)}
\def\etal{{\it et al.}}



\end{document}